
\newcount\fcount \fcount=0
\def\ref#1{\global\advance\fcount by 1 \global\xdef#1{\relax\the\fcount}}

\def\la{\lower.5ex\hbox{$\; \buildrel < \over \sim \;$}}
\def\ga{\lower.5ex\hbox{$\; \buildrel > \over \sim \;$}}

\magnification=\magstep1

\raggedbottom
\footline={\ifnum\pageno=1 \hfil \else \hss\tenrm\folio\hss\fi}

\tolerance = 30000

\baselineskip=5.2mm plus 0.1mm minus 0.1mm
\centerline {\bf CONSTRAINTS ON RADIATIVE DECAY OF THE 17-keV NEUTRINO}
\centerline {\bf FROM COBE MEASUREMENTS}
\vskip 3cm
\centerline {\bf Biman B. Nath}
\vskip 0.5cm
\centerline{Department of Astronomy}
\centerline{University of Maryland}
\centerline{College Park, MD 20742}
\vskip 5cm
\centerline{\bf Abstract}
\bigskip
It is shown that, for a non-trivial radiative decay channel of
the 17-keV neutrino, the photons would distort the microwave background
radiation through ionization of the universe. The constraint on the
branching ratio
of such decays from COBE measurements is found to be more stringent
than that from SN 1987A. The limit on the branching ratio in terms
of the Compton $y$ parameter is $B_{\gamma} \le 1.5\times 10^{-7}
({\tau_{\nu} \over 10^{11}s})^{0.45} \> ({y\over 10^{-3}})^{1.11}\> h^{-1}$
 for an $\Omega=1$, $\Omega_B=0.1$ universe.

\vfill\eject

\noindent
{\bf Introduction}
\bigskip
The existence of a neutrino with a mass of 17 keV and a mixing angle
with $\nu_e$ of about 10\% has been recently suggested from the studies
of $\beta-$decay spectra of different substances\ref\groupa
$^{\groupa}$. Various
constraints have been put on its properties from accelerator data, and
 cosmological and astrophysical considerations\ref\kolbt
$^{\kolbt}$. A number of
theoretical models have also been put forward to accomodate the 17-keV
neutrino and the constraints\ref\groupb$^{\groupb}$.

In brief, constraints from neutrinoless $\beta\beta$-decay and upper limits on
neutrino mass seem to suggest that it is predominantly a
17-keV $\tau$ neutrino (see, e.g., ref. \kolbt).
To avoid making age of the universe too small, the heavy neutrino must decay
into
relativistic particles, with a $\tau_{\nu_H} \le 4 \times 10^{12} (\Omega
h^2)^{3/2}$ sec, if the comoving number density is not decreased
enormously by some exotic process (entropy production or enhanced
annihilation) after the `freeze-out'.
This implies that, in a radiation dominated universe, the
redshift of decay $z_d \geq 190$. The possible decay channels that
have been discussed
are (a) $\nu_H \rightarrow \nu_L+\gamma$, (b) $\nu_H \rightarrow \nu_L+
\chi$, and (c) $\nu_H \rightarrow 3 \nu_L$, where $\nu_{H,L}$ denote
heavy and light neutrinos and $\chi$ denotes a pseudo-scalar.

Cosmological and astrophysical implications of the radiative decay
are interesting. Observations from the supernova 1987A have been used
to put an upper bound on the branching ratio of the radiative decay,
 $B_{\gamma}$\ref\supernova$^{\supernova}$. The
failure of the Solar Maximum Mission satellite
to detect any $\gamma$-ray signal from 1987A indicates that
$B_{\gamma} \le \tau_{\nu}/ 4.9 \times 10^{13}$ sec.
Dicus, Kolb, and Teplitz\ref\dicus$^{\dicus}$
discussed a limit on the lifetime for radiative decays of
massive neutrinos in a hot and
{\it ionized} universe. They argued that if $B_\gamma\sim 1$, then the
energetic
photons would need enough time to thermalize to the energy density
of the background radiation, which limits the lifetime
to less than $\sim 1$ year. This has been interpreted as a hint that
radiative decay is not the dominant decay mechanism for the massive neutrino.

In the standard cosmology, the universe recombined at around $z \sim
1100$ \ref\book$^{\book}$ . If the heavy neutrinos decayed
 after this epoch, the
energetic photons would first interact with neutral atoms.
The processes of photo-ionization and recombination would then be
important. Collisional ionization by the hot electrons could
 aid the decay photon in ionizing the
universe and the cosmic background radiation (CBR) spectrum would become
vulnerable to distortion by the hot electrons by inverse Compton
scattering\ref\fukugita$^{\fukugita}$. This process, therefore, offers us
another
independent limit on the branching ratio for radiative decay.

In this paper we show that the distortion of the CBR spectrum
 through ionization of the universe and inverse Compton effect is
significant. The recent
measurement of the CBR spectrum from COBE is then shown to put
severe bounds on $B_{\gamma}$ as a function of $\tau_{\nu}$.
\bigskip

\noindent
{\bf Ionization of the universe}
\bigskip

Consider the following scenario. The heavy neutrino decays into relativistic
particles at a redshift $z_d$ along with a fraction $B_\gamma$ of it
decaying into photons. The relativistic particles make the
universe radiation dominated at $z_d$ $^{\kolbt}$.
 For such a universe, the corresponding lifetime
of the heavy neutrino,
$\tau_{\nu H} = {1\over 2H_o (1+z)^2} \sim
{1.5 \times10^{17}sec\> h^{-1}\over (1+z)^2}$. We shall be concerned with
decays after the recombination epoch in the standard big bang model,
i.e., $z_d \leq 1100\> (\tau_{\nu H}\geq 1.24\times 10^{11}sec\> h^{-1})$.
 For simplicity, we shall assume a universe with $\Omega=1$, $\Omega_b=0.1$.
We shall also assume that the decays are instantaneous.

Initially, photo-ionization of the neutral atoms by the decay photons produces
the first hot electrons. Collisional ionization by the energetic electron then
takes over. Recombination of ions and electrons competes
with these ionizing processes and is characterized by $\alpha_{rec}$, the
recombination coefficient. The evolution
of the fraction of ionization, $f(=n_{ion}/n_{neutral})$, is governed by the
equation

$$
n_{total}{df\over dt}=n_{neutral}\>(\sigma_{photo}\>n_{\gamma}c\>
+\sigma_{coll}(T_e)\>n_e(T_e)\>v_e)
-\alpha_{rec}(T_e)n_e^2,\eqno(1)
$$

where $\sigma_{photo}$ and $\sigma_{coll}$ are the ionization cross-sections
 for photons and electrons, $T_e$ is the electron temperature, $v_e$ is the
thermal
velocity, $n_{total}\>=n_{neutral}+n_{ion}$ (number per unit volume), and
we assume that $n_e=n_{ion}$. Recombinations also lead to energetic
photons and further ionizations. If the recombinations to the ground level
are the only ones producing new ionizing photons, and if these
are assumed to cause ionizations only locally\ref\onthespot$^{\onthespot}$,
then recombinations to all but the ground level need to be considered. We shall
use $\alpha_{rec} = 2\times 10^{-11} \>T_e^{-1/2}$ cm$^3$/sec ($T_e$ in the
units
of degree kelvin) and

$$
\sigma_{photo}=5.1\times 10^{-20} cm^2\>({E_{\gamma}\over 250eV})^{-p},\qquad
\eqalign{p&=2.65,\;  25eV \leq E_{\gamma} \leq 250eV \cr
&=3.30. \; E_{\gamma} \geq 250eV\cr} \eqno(2)
$$

The latter is particularly appropriate for a mixture of
helium and hydrogen atoms with primordial abundance\ref\zd
$^{\zd}$. For collisional
ionization, we shall use $\sigma_{coll}=4 \times 10^{-14} cm^2
{ln(U)\over \chi^2 U}$, where $\chi$ is the ionization threshold in eV (13.6
for hydrogen) and $U$ is energy of the electron in the units of $\chi$
\ref\lotz$^{\lotz}$.
 In an $\Omega=1$, $\Omega_B=0.1$ universe, $n_{total}\> (z=0)
\sim 10^{-6} h^2$, assuming primordial abundance and $n_{total}
\propto (1+z)^3$ due to the expansion of the universe.
The number density of neutrinos is given by
the big bang nucleosynthesis model: $n_{\nu}
=({3\over11})n_{\gamma CBR}\sim120$ per cc at $z=0$. The number density
of decay photons $n_\gamma$ is $B_\gamma n_\nu$.

In solving $(1)$,
we should be careful not to use a single photon for more than one ionization
event. In other words, as a photon ionizes an atom, it no longer remains
available for further ionization. We take this into account by using
$n_{\gamma eff}$ =
$f_{\gamma eff}$ $n_{\gamma}$ instead of $n_\gamma$, where

$$
n_\gamma{df_{\gamma eff}\over dt}= - n_{neutral}\>\sigma_{photo}n_{\gamma
eff}c.
\eqno(3)
$$

The first equation is thus rewritten as

$$
n_{total}{df\over dt}=n_{neutral}\>(\sigma_{photo}n_{\gamma eff}c
+\sigma_{coll}(T_e)\>n_e(T_e)\>v_e)
-\alpha_{rec}(T_e)n_e^2.\eqno(1a)
$$

The energy density in electrons $\epsilon_e (={3\over2}n_ekT_e)$
is to be determined by considering various energy sources and sinks.
Recombinations and inverse Compton scattering of
CBR photons reduce the electron energy density
$\epsilon_e$, whereas photo-ionization of newly
recombined and remaining neutral atoms adds to it. Compton scattering
by the decay photons transfers energy to the electrons.
Adiabatic expansion of the universe becomes important only
later, at smaller redshifts. The energy equation is the following:

$$
{d\epsilon_e\over dt}=-({\epsilon_e\over t_{i.C.}})-({\epsilon_e\over
t_{rec}}) - ({5\epsilon_e\over 2 t_{U}})
+(E_{photo}\>n_{neutral}\>\sigma_{photo}+
{E_\gamma-4kT_e\over m_ec^2}\>E_\gamma\>n_e\>\sigma_T)
\>n_{\gamma eff}c.\eqno(4)
$$

Here $E_{photo}$
is the energy gained by a photo-electron from the decay photon, viz.
$h\nu- 13.6eV$ (for photo-ionization of a hydrogen atom);
 $t_{i.C.}$ and $t_{rec}$ are the timescales for inverse
Compton scattering and recombination, and $t_U$ is the age of the universe.
The last term in the equation denotes the energy input by Compton
scattering. The energy of the decay photon is $\sim {m_{\nu H}
\over2}{(1+z)\over(1+z_d)}$.
Inverse Compton scattering transfers energy to the CBR in a timescale
$t_{i.C.}={m_ec^2\gamma\over {4\over3}\sigma_Tc\gamma^2\beta^2aT_{CBR}^4}
\sim {2.6\times 10^{29} sec \over T_e (1+z)^4}$, for $T_{CBR}(z=0)=
2.75^o\>K$.
The recombination timescale, $t_{rec}={1\over n_e \alpha_{rec}(T_e)}
\sim 5\times 10^{12} sec\> ({n_e\over 1/cc})^{-1} ({T_e\over 10^4})^{1/2}$.
In a radiation dominated universe, the age of the universe $t_U\>\sim
{1.5 \times10^{17}sec\> h^{-1}\over (1+z)^2}$.

The three equations $(1a)$, $ (3)$ and $(4)$, describing the number
densities of ionized matter and decay photons, and the electron energy
density, can be numerically solved for specific
values of $z_d$ and $B_{\gamma}$. Fig. 1 shows the evolution of
$\epsilon_e$ for a few cases. Fig. 2 shows the evolution of electron
temperature
$T_e$. Initially, photo-ionization
deposits energy to the electrons till collisional ionization takes over
and energy input is then due only to Compton scattering.
At high redshift Compton heating competes mainly with
cooling due to recombination.
Later, as the energy of the decay photons
decreases, photo-ionization becomes important again and deposits
more energy in the electrons. Thus, photo-ionization is important
only near the decay redshift, when the first ionizations occur, and
at smaller redshifts, when the cross-section has increased and
an efficient recombination process (due to lowered temperature)
produces new neutral atoms to be ionized. For bigger $B_{\gamma}$,
 the initial temperature is large and $T_e$ never drops below
$10^{5} $ K for effective recombination and a second phase of
photo-ionization to occur.
\bigskip
\noindent
{\bf Limits from COBE}
\bigskip

Armed with the knowledge of $\epsilon_e$ as a function of time, we can now
calculate the distortion of the CBR. The distortion in terms of
the Compton $y$ parameter, characterizing the deviation of the spectrum
from Planckian, is given by
$$
{dy\over dt}={\epsilon_e\over m_ec^2} \sigma_{T} c.\eqno(5)
$$
Fig. 3 shows the resulting $y$ for various $B_\gamma$ as a function of
$z_d$.

The current upper bound on $y$ comes from COBE and is $\sim 10^{-3}$
\ref\cobe$^{\cobe}$.
This limit can be translated into a bound on $B_\gamma$ as a function of
$\tau_\nu$. The
limit from SN 1987A is shown along with the COBE limit in Fig. 4. The limit
that COBE may give soon ($y < 10^{-4}$), with the collection
of new data and continued analysis, is also indicated. The limiting curves in
Fig. 4 are well fitted by $B_{\gamma} \le 1.5 \times 10^{-7}
({\tau_{\nu} \over 10^{11}s})^{0.45}\> ({y\over 10^{-3}})^{1.11}\>h^{-1}$.

Upper bounds for $B_{\gamma}$ have also been sought in the past from
the diffuse $\gamma$-ray background flux. McKeller and Pakvasa\ref\pakvasa
$^{\pakvasa}$ found that, for $m_{\nu H}\sim 75$ keV and $\tau_{\nu H}\sim
10^{11}$ sec, $B_{\gamma}\le 1.5\times10^{-6}$. The limit from COBE is
certainly stronger than this.

A note on the difference between our result and that of Altherr
{et. al}'s recent paper\ref\altherr$^{\altherr}$ is in order here. They
estimated a bound on $B_{\gamma}$ from the COBE limit on the
chemical potential, $\mu$, of the microwave background.
Their result shows that the bound calculated above from Compton
$y$ parameter is stronger than that from $\mu$ for decays
after recombination. The bound is going to be even stronger
if COBE limit on $y$ goes down to $10^{-4}$.

Our result pertains only to the process of photoionization of the universe
and subsequent distortion of the microwave background. We, therefore,
have not considered the case of decays before recombination;
the interaction of photons with baryons in that case would be
different from the scenario sketched above.
It is hoped that the bound on $B_{\gamma}$
calculated above will be useful in constructing theoretical
models for the massive neutrino.

\bigskip
\centerline{\bf Acknowledgements}
\medskip
I am indebted to Dr. David Eichler for encouraging me and for his
valuable comments. I thank Drs. Rabindra Mohapatra, David Spergel
and Virginia Trimble for their comments on the manuscript. I am also
grateful to Ravi  Kuchimanchi and Bikram Phookun, who kept me
stimulated with many grueling questions.

\vfill\eject

\centerline{\bf References}
\def\refs{\hangindent=5ex\hangafter=1}
\parskip=0pt
\parindent=0pt

\medskip

\refs\groupa . J. Simpson, Phys. Rev. Lett. {\bf54}, 1891 (1985);
 J. Simpson and A. Hime, Phys. Rev. D {\bf39}, 1825 (1989); {\bf39},
 1837 (1989); A. Hime and N. A. Jelley, Phys. Lett. B {\bf257},
 441 (1991).

\refs\kolbt . E. Kolb and M. Turner, Phys. Rev. Lett.,
 {\bf67}, 5 (1991).

\refs\groupb . S. L. Glashow, Phys. Lett. B {\bf256}, 218 (1991);
 K. S. Babu, R. N. Mohapatra, and I. Z. Rothstein, Phys. Rev. Lett.,
{\bf67}, 545 (1991).

\refs\supernova . E. Kolb and M. Turner, Phys. Rev. Lett.,
 {\bf62}, 509 (1987).

\refs\dicus . D. A. Dicus, E. W. Kolb, and V. L. Teplitz, Astrophys.
J., {\bf221}, 327 (1978).

\refs\book . E. Kolb. and M. Turner, The Early Universe,
(Addison-Wesley, 1989).

\refs\fukugita . Mechanism of a similar kind to distort the CBR spectrum
was pursued a few years ago to explain the spurious measurement of a bump in
the
spectrum which was later proven wrong by COBE (see, e.g., M. Fukugita, Phys.
Rev. Lett., {\bf61}, 1046 (1988) and the rebuttal, e.g., by G. Field and T.
Walker,
Phys. Rev. Lett., {\bf63}, 117 (1989)).

\refs\onthespot . When recombination is appreciable, i.e., at $T_e \sim 10^5$
K,
the mean free time of an ionizing photon is $\sim {10^{13}h^{-2}\over(1+z)^3}$
sec, many orders of magnitude smaller than the characteristic expansion time
$t_U \> (\sim{10^{17} h^{-1}\over (1+z)^2}$sec) of the universe.

\refs\zd . A. A. Zdziarski and R. Svensson, Astrophys. J., {\bf344},
 551 (1989).

\refs\lotz . W. Lotz, Astrophys. J. Suppl., {\bf14}, 207 (1967).

\refs\cobe . J. C. Mather {\it et al.}, Ap. J. Lett.,
        {\bf354}, L37-L40 (1990).

\refs\pakvasa . B. H. J. McKeller and S. Pakvasa, Phys. Lett.,
{\bf 122B}, 33 (1983).

\refs\altherr . T. Altherr, P. Chardonnet and P. Salati, Phys. Lett.,
{\bf 265B}, 251 (1991).

\vfill\eject
\centerline{\bf Figure Captions}

Figure 1. Evolution of electron energy density with
redshift for $1+ z_d= 200, 500$ and $B_{\gamma}=10^{-6}, 10^{-7}$.

Figure 2. Evolution of electron temperature with redshift for $1+z_d=
200, 500$ and $B_{\gamma}=10^{-7}$.

Figure 3. Distortion of the CBR spectrum for various branching
ratios as a function of the decay redshift.

Figure 4. Limiting curve for the branching ratio from COBE is shown
along with the limit from SN 1987A.
The dotted and solid curves correspond to the cases $h=$0.5 and 1.0
respectively. The lowest set of curves denote
the limit that COBE may give soon with $y\le 10^{-4}$.
\end